\documentclass[aps,prl,groupedaddress,10pt,twocolumn,floatfix]{revtex4-1}
\usepackage{bbm}
\usepackage[final]{graphics}
\usepackage{amsmath}
\usepackage{amsfonts,amsbsy}
\usepackage{amssymb}
\usepackage{bm}
\usepackage{color}

\def\empile#1\over#2{\mathrel{\mathop{\kern 0pt#1}\limits_{#2}}}

\def\TODO#1{}

\def\k{{\boldsymbol k}}

\def\x{{\boldsymbol x}}
\def\y{{\boldsymbol y}}

\begin{document}
\title{Pressure isotropization in high energy heavy ion collisions}
\author{Thomas Epelbaum}
\email{thomas.epelbaum@cea.fr}
\author{Fran\c cois Gelis}
\email{francois.gelis@cea.fr}
\affiliation{Institut de Physique Th\'eorique, CEA/Saclay, 91191 Gif sur Yvette cedex, France}

\date{\today}

\begin{abstract}
  The early stages of high energy heavy ion collisions are studied in
  the Color Glass Condensate framework, with a real-time classical
  lattice simulation. When increasing the coupling constant, we
  observe a rapid increase of the ratio of longitudinal to transverse
  pressure. The transient regime that precedes this behavior is of the order of $1$ fm/c.
\end{abstract}


\maketitle

\section{Introduction}
Heavy ion collisions at ultrarelativistic energies are currently being
performed at the Relativistic Heavy Ion Collider (RHIC) and the Large
Hadron Collider (LHC), in order to study the properties of nuclear
matter at extreme temperatures and densities. Models that assume that the fireball
produced in these collisions expands according to
the laws of relativistic hydrodynamics have been very successful in
reproducing the behavior of many bulk
observables~\cite{Teane1,Teane2,Ollit2,OllitG1}.

Hydrodynamical flow follows from energy-momentum
conservation, but also assumes that the energy-momentum tensor of
the system is sufficiently close to that of a
perfect fluid, that reads
\begin{equation}
T^{\mu\nu}_{\rm perfect}
={\rm diag}\,(\epsilon,p,p,p)\; ,
\label{eq:Tmunu-ideal}
\end{equation}
where $\epsilon$ is the energy density and $p$ the pressure (related
to $\epsilon$ by an equation of state). In particular, the pressure
tensor of a perfect fluid at rest is isotropic. A limited amount of
pressure anisotropy can be accommodated in the hydrodynamical
description by adding to eq.~(\ref{eq:Tmunu-ideal}) some viscous
corrections.

Understanding why the pressure tensor becomes nearly isotropic in
terms of the underlying Quantum-Chromodynamics (QCD) dynamics has so
far been very challenging~\cite{BergeBG1}. At high energy, the density of constituents
(mostly gluons) in the two nuclei is large, and a consistent QCD
description can be achieved in the Color Glass Condensate (CGC)
framework~\cite{Lappi6,GelisIJV1}. The CGC is designed to collect and
sum all the recombination and multiple scattering corrections that are
prevalent at high gluon density. These nonlinear effects are
controlled by a unique dimensionful parameter, the saturation momentum
$Q_s$, that increases with energy.

For inclusive quantities, like the expectation value of the
energy-momentum tensor, the CGC provides an expansion in powers of the
strong coupling constant $\alpha_s=\frac{g^{2}}{4\pi}$, in which the leading order (LO)
is obtained by solving the classical
Yang-Mills equations. Immediately after the collision (proper time
$\tau=0^+$, see the figure \ref{fig:setup1}), the CGC gives the
following energy-momentum tensor at LO~\cite{LappiM1},
\begin{equation}
T^{\mu\nu}_{_{\rm CGC,LO}} = {\rm diag}\,(\epsilon,\epsilon,\epsilon,-\epsilon)\; .
\end{equation}
The longitudinal
pressure is negative, exactly opposite to the energy density and the
transverse pressure, and therefore quite far from the near ideal form
expected when hydrodynamics is applicable. 

However, for anisotropic systems, classical
solutions of the Yang-Mills equations are subject to Weibel
instabilities that make them exponentially sensitive to their initial
conditions~\cite{RomatV1,RomatV3}. These instabilities are triggered
by next-to-leading order (NLO) corrections, in which they cause
secular divergences, i.e. terms that are of higher order in
the coupling $\alpha_s$ but accompanied by a coefficient that becomes
infinite when $\tau\to+\infty$.  These secular terms break the simple power counting that organizes the expansion
in power of $\alpha_s$. It is however possible to resum at all orders
in $\alpha_s$ the terms that grow the fastest in
time~\cite{GelisLV3,DusliGV1}, by allowing the initial fields to fluctuate with a Gaussian distribution whose
variance is given by a 1-loop calculation.

This resummed result includes the LO and NLO
contributions, plus a subset of all the higher orders, and it remains
finite at all times. Moreover, it has been shown in the case of a
scalar theory that this reorganization of the perturbative expansion
leads to the near isotropy of the pressure tensor, and to a good
agreement with nearly ideal hydrodynamics~\cite{DusliEGV2}.  The
purpose of this paper is to study this
resummation in the case of Yang-Mills theory, which is directly
relevant for heavy ion collisions. We use the
classical statistical approach --performing a Monte-Carlo
sampling of the Gaussian ensemble of classical initial conditions and
solving numerically the classical Yang-Mills equations in real time
on a 3 dimensional lattice-- in order to study the time dependence
of the energy-momentum tensor shortly after the collision.

\section{Spectrum of initial conditions} 
In the absence of quantum fluctuations (i.e. at LO), the initial gauge fields and
electrical fields are given by classical solutions of the Yang-Mills
equations in the presence of two light-cone color currents representing the two nuclei. At $\tau=0^+$, these solutions read
\begin{eqnarray}
&
A_0^i=\alpha_1^i+\alpha_2^i\quad,\;
&
E_0^i=0\quad,\;
\alpha_n^i=\frac{i}{g}U_n^\dagger \partial^i U_n\; ,
\nonumber\\
&
A_{0\eta}=0\quad,\;
&
E_0^\eta=i\frac{g}{2}[\alpha_1^i,\alpha_2^i]\;,
\label{eq:bkg}
\end{eqnarray}
where the Wilson lines $U_{1,2}(\x_\perp)$ read
\begin{equation}
U_1(\x_\perp)
=
{\rm T}\,e^{ig\int dx^- \frac{1}{\nabla_\perp^2}\rho_1(x^-,\x_\perp)}
\end{equation}
in the McLerran-Venugopalan model~\cite{McLerV2}.
 The saturation momentum $Q_s$ controls the event-by-event fluctuations
of the color charge density $\rho_{1}$
\begin{equation}
g^2\int dx^- dy^-\,\left<\rho_1^a(x)\rho_1^b(y)\right>
=
\delta^{ab}Q_s^2\delta(\x_\perp-\y_\perp)\; .
\end{equation}
The $U_2$ of the second nucleus is obtained similarly. To this background field, we add a fluctuating component, to obtain
\begin{eqnarray}
A^\mu(x) &=& A_0^\mu(x) \!+\! \sum_{c,\lambda}\!\int\! \frac{d^3\k}{(2\pi)^3 2k} 
\;\Big[c_{c\lambda\k}\,a^\mu_{c\lambda\k}(x)+\mbox{\rm c.c.}\Big]
\nonumber\\
E^\mu(x) &=& E_0^\mu(x) \!+ \!\sum_{c,\lambda}\!\int \!\frac{d^3\k}{(2\pi)^3 2k} 
\;\Big[c_{c\lambda\k}\,e^\mu_{c\lambda\k}(x)+\mbox{\rm c.c.}\Big]\, ,
\label{eq:fluct}
\end{eqnarray}
where $a^\mu_{c\lambda\k}(x)$ ($e^\mu_{c\lambda\k}(x)$ is the
conjugate electrical field) is the solution of the linearized
Yang-Mills equations over the background field $A_0^\mu$, with as
initial condition at $x^0=-\infty$ a plane wave of color $c$,
polarization $\lambda$ and momentum $\k$. The coefficients
$c_{c\lambda\k}$ are complex Gaussian distributed random numbers,
whose variance is
\begin{equation}
  \left<c_{c\lambda\k}c_{c'\lambda'\k'}^*\right>=(2\pi)^3 k \;
\delta_{cc'}\delta_{\lambda\lambda'}\delta(\k-\k')\; .
\end{equation}
Note that the background field is of order $Q_s/g$ while the
fluctuating part is of order $Q_s$.

The mode functions $a^\mu_{c\lambda\k}(x), e^\mu_{c\lambda\k}(x)$ have
been determined analytically in Ref.~\cite{EpelbG2} (see the Eqs.~(69)
in this reference), in terms of formulas involving only Fourier
integrals, at a time $\tau_0\ll Q_s^{-1}$ (i.e. just after the
collision). Numerically, we proceed as follows:
\begin{itemize}
\item[{\bf i.}] Compute the background fields according to
  Eqs.~(\ref{eq:bkg}).  Since we are interested in studying
  isotropization in a given event, a single configuration of this
  background field is generated.
\item[{\bf ii.}] Generate random Gaussian numbers $c_{c\lambda\k}$,
  and evaluate Eqs.~(\ref{eq:fluct}) at some small initial time
  $\tau_0\ll Q_s^{-1}$,
\item[{\bf iii.}] Using this configuration $A^\mu,E^\mu$ as initial
  condition at $\tau_0$, solve the classical Yang-Mills equations
  $\partial_0 A^\mu = E^\mu\;,\; \partial_0 E^\mu = D_i F^{i\mu}+D_\eta F^{\eta\mu}$
  (written here in temporal gauge $A^0=0$) up to the largest time
  of interest,
\item[{\bf iv.}] Evaluate the observable in terms of this classical solution,
\item[{\bf v.}] Repeat the steps {\bf ii}--{\bf iv} in order to sample
  the ensemble of fluctuating initial conditions.
\end{itemize}

\section{Simulation setup}
\begin{figure}[htbp]
\resizebox*{4.5cm}{!}{\includegraphics{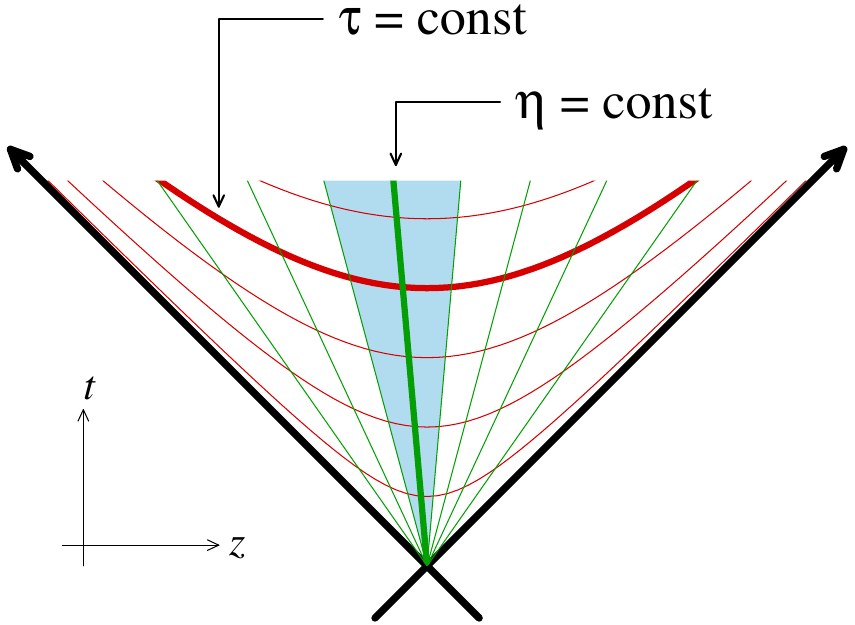}}
\caption{\label{fig:setup1}Comoving coordinate system in the forward
  light-cone of the collision point.}
\end{figure}
In order to cope with the longitudinal expansion of the system, we use
the comoving coordinates $\tau\equiv \sqrt{t^2-z^2}$ (proper time) and
$ \eta\equiv \frac{1}{2}\ln(t+z)/(t-z)$ (rapidity). As illustrated in
the figure \ref{fig:setup1}, a constant extent in rapidity corresponds
to a volume that expands in the longitudinal direction as time
increases.

\begin{figure}[htbp]
\resizebox*{8.6cm}{!}{\includegraphics{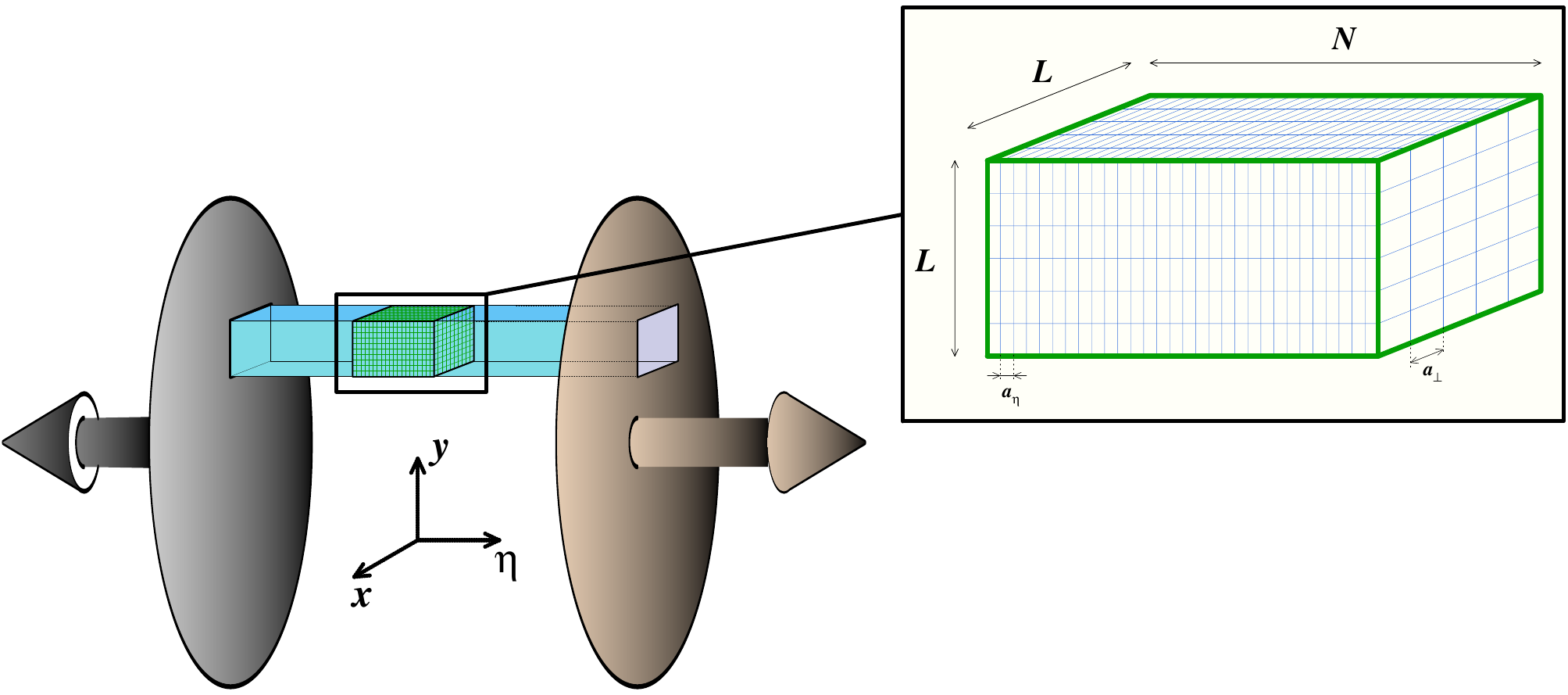}}
\caption{\label{fig:setup}Lattice setup.}
\end{figure}

We solve the Yang-Mills equations numerically by discretizing space,
while time remains a continuous variable whose increments can be
arbitrarily small, as needed to ensure the accuracy of the solution.
Due to limited computational resources, we do not simulate the entire
interaction zone, but only a smaller sub-volume, both in the transverse
directions and in rapidity (see the figure \ref{fig:setup}).
For reasons related to the longitudinal expansion of the system, it is
necessary to have a larger number ($N$) of lattice intervals in the
longitudinal direction than in the transverse ones ($L$).

The results presented in this paper were obtained on a
$64\times64\times128$ lattice with $N_{\rm conf}$ field configurations in the 
Monte-Carlo sampling. The lattice spacing in the rapidity
direction is set to $a_\eta=1/64$, so that our lattice covers two
units of rapidity. The saturation momentum $Q_s$ was chosen such that
$Q_s a_\perp=1$, i.e. significantly below the lattice ultraviolet
cutoff for transverse momenta ($k_\perp^{\rm max}a_\perp=\sqrt{8}$). A study of the dependence on the lattice parameters, that may affect our results, will be performed in a future work. The field configurations are generated at the
initial time $Q_s\tau_0=0.01$, but the subsequent results do not
depend on this choice as long as $Q_s\tau_0\ll 1$. In order to
simplify the color algebra, the simulation is done for an SU(2) gauge
group, instead of SU(3) for actual QCD.

We work in Fock-Schwinger gauge, $A^\tau=0$, that generalizes the
temporal gauge to the $(\tau,\eta,\x_\perp)$ system of coordinates,
and has the advantage of treating the two nuclei on the same
footing. On the lattice,
the vector potentials are exponentiated into link variables that
connect adjacent lattice sites in order to preserve an exact local
gauge symmetry. However, exponentiating the vector potentials in
Eqs.~(\ref{eq:fluct}) introduces some small violations of Gauss's law
$D_\mu E^\mu=0$. We restore Gauss's law by projecting the initial electrical
fields on the subspace that obeys the constraint, using the algorithm
described in Ref.~\cite{Moore3}.

\section{Energy-momentum tensor}
From the solutions of the classical Yang-Mills equations at some time
$\tau$, we compute the expectation value of the components of the
energy momentum tensor.  In order to increase the effective
statistics, we average both over the random numbers $c_{\nu\lambda\k}$
of Eqs.~(\ref{eq:fluct}) and over the lattice volume.  At all times,
the transverse and longitudinal pressures are related to the energy
density by $\epsilon=2P_{_T}+P_{_L}$ (by construction), and Bjorken's
law,
\begin{equation}
\frac{\partial\epsilon}{\partial\tau}+\frac{\epsilon+P_{_L}}{\tau}=0\; ,
\label{eq:BL}
\end{equation}
is satisfied as a consequence of energy and momentum conservation.

The energy-momentum tensor computed in this approach contains a zero point
 contribution, that exists even when the background field in
Eqs.~(\ref{eq:bkg}) is set to zero. We subtract it out by performing the same calculation
twice: with a background field generated with a non-zero value of
$Q_s$ and with the background field set to zero. 

After this pure vacuum subtraction, $\epsilon$ and $P_{_L}$ still contain subleading
divergences that behave as $\tau^{-2}$. Although we cannot compute the
corresponding counterterms from first principles at the moment, their
form can be predicted. From $\epsilon=2 P_{_T}+P_{_L}$, the
counterterms for $\epsilon$ and $P_{_L}$ must be equal. Then Bjorken's
law (\ref{eq:BL}) constrains this common counterterm to be of the form
$A/\tau^2$. We fit the prefactor $A$ in order to make $\epsilon$ and
$P_{_L}$ finite in the limit $\tau\to 0^+$. This choice of $A$ also
makes the resummed and Leading Order results very close when
$\tau\to0^+$, which is expected since the higher order corrections
should be important only at later times, after the fluctuations have been
amplified by the Weibel instability.

To summarize our procedure, we do
\begin{equation}
\begin{aligned}
\left<P_{_T}\right>_{{\rm phys.}}
&=
\left<P_{_T}\right>_{{\rm backgd.}\atop{\rm+\ fluct.}}
&-
\left<P_{_T}\right>_{{\rm fluct.}\atop{\rm only}}
&
\\
\left<\epsilon,P_{_L}\right>_{{\rm phys.}}
&
=
\underbrace{\left<\epsilon,P_{_L}\right>_{{\rm backgd.}\atop{\rm+\ fluct.}}}_{\rm computed}
&-
\underbrace{\left<\epsilon,P_{_L}\right>_{{\rm fluct.}\atop{\rm only}}}_{\rm computed}
&+
\underbrace{A\,\tau^{-2}\vphantom{\left<\epsilon,P_{_L}\right>_{{\rm fluct.}\atop{\rm only}}}}_{\rm fitted}
\; .
\end{aligned}
\end{equation}

\begin{figure}[htbp]
\resizebox*{8.6cm}{!}{\includegraphics{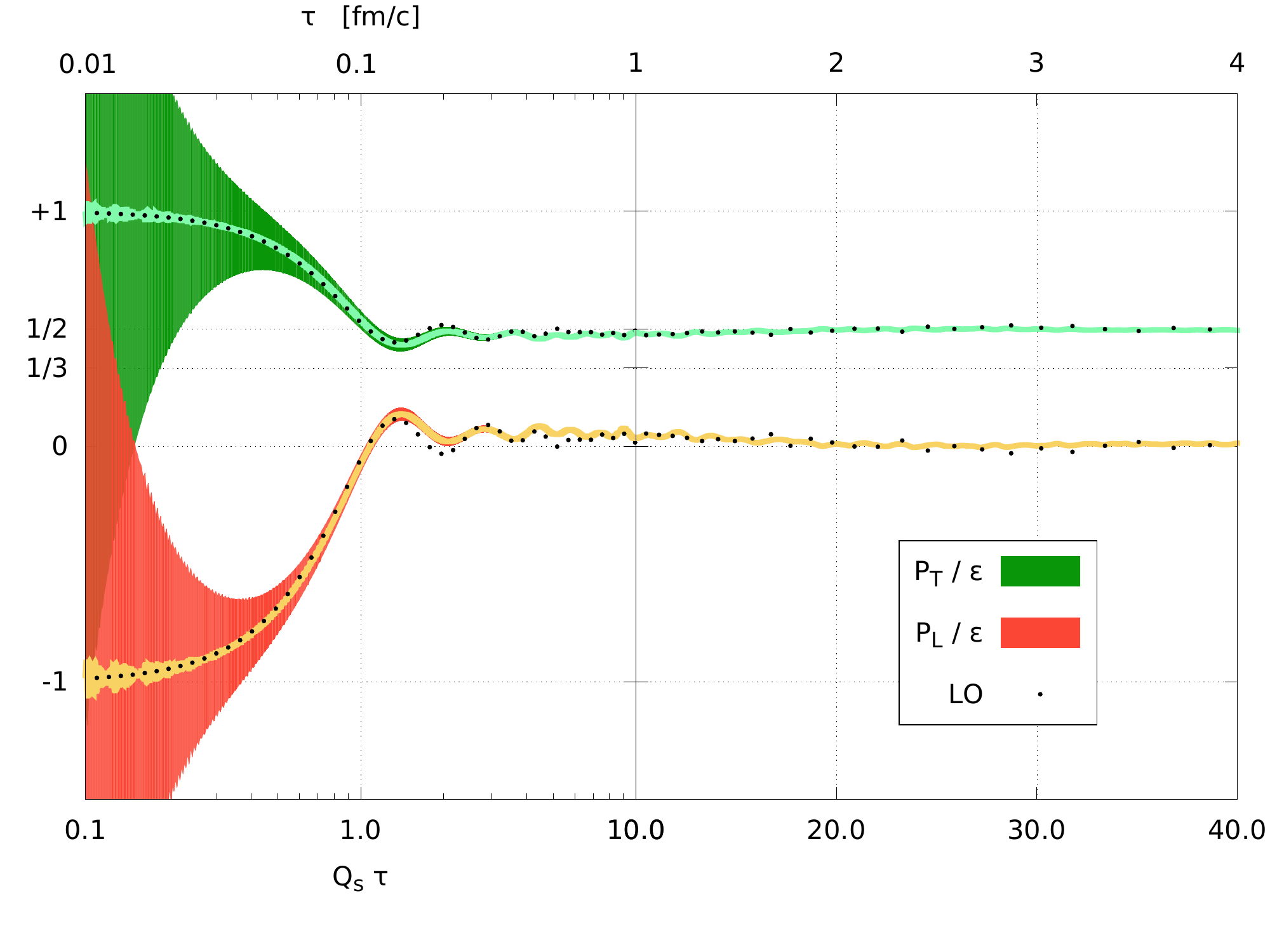}}
\caption{\label{fig:g0.1}Evolution of the ratios $P_{_{T,L}}/\epsilon$
  for $g=0.1$ ($\alpha_s=8\cdot10^{-4}$). The bands indicate
  statistical errors, estimated as the result obtained before any subtraction divided by the square root of the number of samples. The dotted curves represent the LO result.}
\end{figure}

\begin{figure}[htbp]
\resizebox*{8.6cm}{!}{\includegraphics{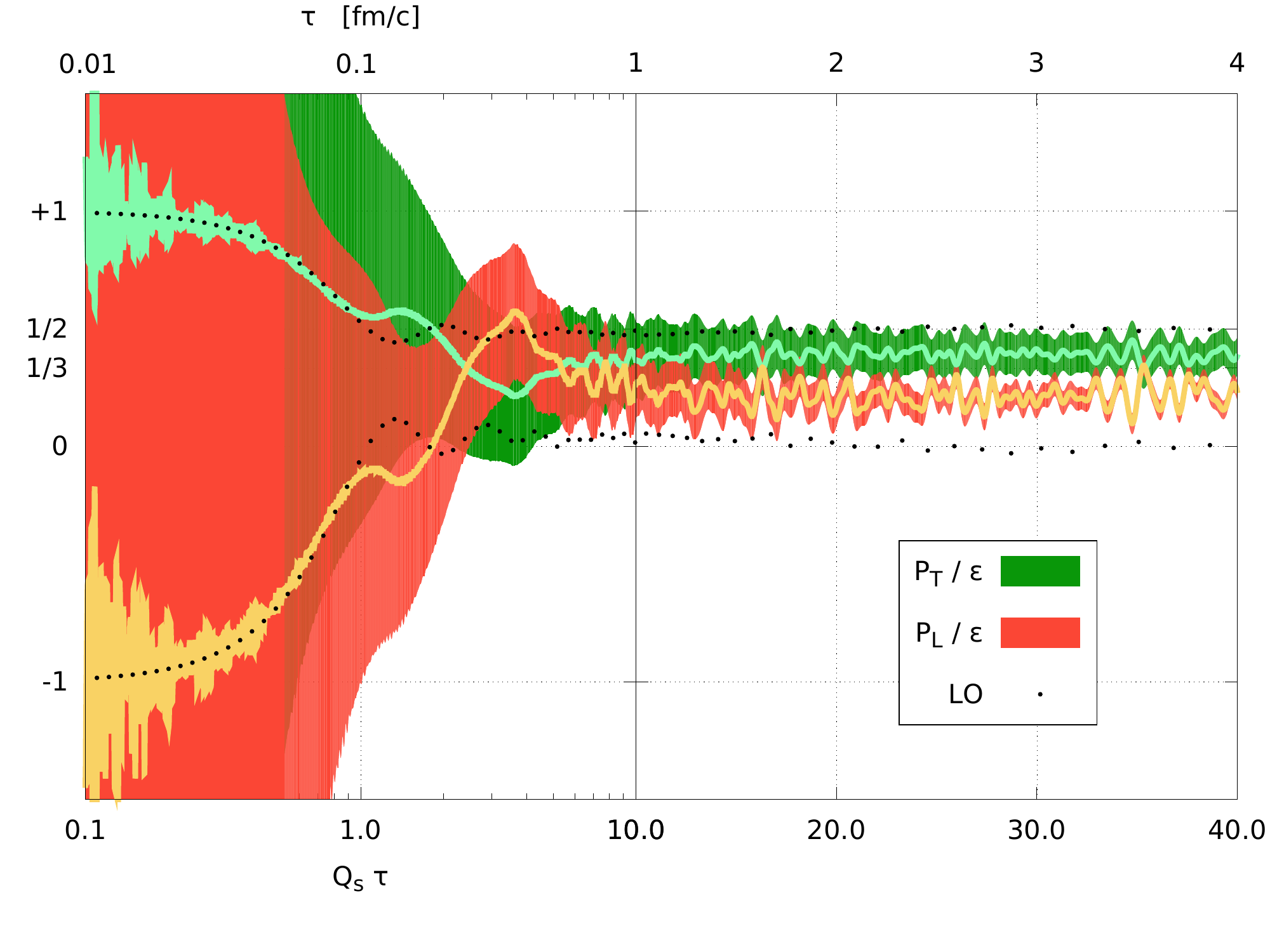}}
\caption{\label{fig:g0.5}Evolution of the ratios $P_{_{T,L}}/\epsilon$
  for $g=0.5$ ($\alpha_s=2\cdot 10^{-2}$).}
\end{figure}

It should be noted that the zero point contribution also behaves as
$\tau^{-2}$ at small times and is almost independent of the coupling,
while the physical contribution is of order $Q_s^4/g^2$. At large
coupling and small times, the physical contribution is much smaller
than the two terms that we must subtract, and therefore the accuracy
on the difference is severely limited by the statistical errors. 
This limits how large the coupling constant $g$
can be in practical simulations. The results presented below are for
$g=0.1$ (figure \ref{fig:g0.1}, $N_{\rm conf}=200$) and $g=0.5$
(figure \ref{fig:g0.5}, $N_{\rm conf}=2000$), that are
both much smaller than the expected value at the LHC ($g\approx 2$).

To provide more intuition on the relevant timescales, we also provide
the time in fermis/c on the upper horizontal scale of the figures
\ref{fig:g0.1} and \ref{fig:g0.5}. The calibration of this scale
requires that one chooses the value of $Q_s$ in GeV, here taken to be
$Q_s=2~$GeV, a reasonable value for nucleus-nucleus collisions at LHC
energies. In order to highlight the effect of the quantum corrections, we
also show the Leading Order results (dotted curves). 

In both cases, $\epsilon=P_T=-P_L$ at $\tau=0^+$. 
After a time of order $Q_s^{-1}$, the longitudinal pressure
turns positive and stays mostly positive afterwards. However, for
$g=0.1$ it always stays much smaller than the
transverse pressure ($P_{_L}/P_{_T}\approx 0.01$), which implies that
the system is almost free streaming in the longitudinal direction: the
energy density decreases approximately as $\tau^{-1}$. Moreover, 
the result are always very close to the LO results,
suggesting that at such small couplings the Weibel instability does
not play an important role over the timescales we have considered.

Even though $g=0.5$ is still a very weak coupling in QCD, there is
drastic increase in the ratio $P_{_L}/P_{_T}$, that now approaches
$0.60$ at times of the order of one fm/c.  As a consequence, $\epsilon$
falls faster than in free streaming because of the
energy reduction due to the work done by the longitudinal
pressure. Such a degree of residual anisotropy can easily be coped
with in hydrodynamics with moderate viscous corrections. Note that the
pressures have some residual oscillations on shorter timescales of
order $Q_s^{-1}$, that do not affect the long time dynamics. 
The comparison with the LO result now indicates sizable
deviations when $Q_s\tau \gtrsim 1$. Note that the LO results are
identical for $g=0.1$ and $g=0.5$, since at this order the
energy-momentum tensor is given by purely classical field
configurations, from which the coupling dependence can be entirely
factored out.

We have also fitted the time dependence of $\epsilon$ for
$g=0.5$ by assuming that it is governed by hydrodynamics including the
first correction due to shear viscosity,
$\epsilon={a}/{\tau^{4/3}}-2{\eta_0}/{\tau^2}$. This gives an estimate
of the shear viscosity $\eta = \eta_0/\tau$ from which we obtain the
dimensionless ratio $\eta/\epsilon^{3/4}\approx\eta_0/a^{3/4} \approx
0.3$, which is much smaller than the LO perturbation theory value, of
order $\sim 300$ for $g=0.5$ (see Ref.~\cite{ArnolMY1} for $\eta$.
 For $\epsilon$, we use the
Stefan-Boltzmann formula). This is possibly a manifestation of the
anomalously small viscosity conjectured in Ref.~\cite{AsakaBM1} for
systems made of strong disordered fields.

Although the figure \ref{fig:g0.5}, that exhibits isotropization, was
obtained for a coupling which is still much smaller than the
$\alpha_s\approx 0.3$ (i.e. $g\approx 2$) that is expected at the LHC,
we do not expect important qualitative modifications by going at
larger coupling. Moreover, the timescales should not vary much either
(and if anything, one would expect them to become smaller) since the
$g$ dependence is to a large extent cancelled by the fact that the
background fields behave as $g^{-1}$.

\section{Conclusions and outlook}
In this paper, we have presented the first NLO-resummed results in the
Color Glass Condensate framework for the energy-momentum tensor
shortly after a heavy ion collision. At very small coupling, the
system settles on a free-streaming expansion curve, which is not
compatible with ideal hydrodynamics. 

However, by increasing the coupling constant, one reaches a regime of
viscous hydrodynamical expansion, after a fairly short transient
regime that lasts about 1~fm/c for realistic values of the saturation
momentum. This hydrodynamical
regime sets in for very small values of the coupling constant ($g=0.5,
\alpha_s=2\cdot10^{-2}$ in the plots presented above). Although it was
not technically feasible to have a more realistic value
of $g$, we conjecture a similar behavior at larger $g$. Conversely,
the experimental evidence for hydrodynamical flow in heavy ion
collisions does not necessarily imply that the system is strongly
coupled, since weak coupling techniques and resummations already
predict such a behavior.

More systematic studies are necessary in order to assess how one goes
from free streaming at very weak coupling to hydrodynamical behavior
for larger couplings. Moreover, the present study does not tell
how far the system is from local thermal equilibrium when the
hydrodynamical behavior starts. It would be highly interesting to
compute observables that can provide informations on this question.
Recent works, such as
Refs.~\cite{BlaizGLMV1,EpelbG1,BlaizLM2,BergeS4,BergeSS3,KurkeM3},
have investigated the possibility of the formation of Bose-Einstein
condensate when starting from a CGC-like initial condition, since such
a state is overpopulated. It would definitely be important to assess
this question in the present framework. Another important issue
is to develop a rigorous procedure for the
subtractions that we have performed by hand in order to obtain a
finite energy-momentum tensor at short times. Moreover, higher order quantum 
corrections not included here are expected to become important at
late times. Including them is beyond the scope of classical statistical
methods, but at the small couplings we have considered we do not expect
them to be important at the times relevant for pressure isotropization.

This study is related to other recent works on the effect of
instabilities on the early time behavior in heavy ion collisions, in
particular
Refs.~\cite{KurkeM2,KurkeM1,AttemRS1,BergeBS1,BergeS6,BergeBSV1}. The
approach we have pursued, where one solves the classical Yang-Mills
equations with fluctuating initial conditions, is very close to that
of Ref.~\cite{BergeBSV1}, but differs from it in the choice of the
ensemble of initial fields.  In the present work, we have used the
analytical solutions (derived in Ref.~\cite{EpelbG2}) for the mode
functions over the CGC classical gauge fields produced in heavy ion
collisions, while Ref.~\cite{BergeBSV1} used vacuum mode functions,
rescaled in order to obtain a prescribed occupation number at a larger
initial time of order $\tau_0\approx 100\, Q_s^{-1}$. In the future,
it would be interesting to see whether the CGC initial conditions
(that have small fluctuations around a large coherent field) used in
the present paper eventually evolve into the ensemble of fields (that
have no coherent field and large fluctuations) used as the starting
point in Ref.~\cite{BergeBSV1}.

\section*{Acknowledgements}
We thank J. Berges, W. Broniowski, J.-P. Blaizot, L. McLerran,
J.-Y. Ollitrault, S. Schlichting, R. Venugopalan and B. Wu for
discussions related to this work.  This work is supported by the
Agence Nationale de la Recherche project 11-BS04-015-01. Some of the
computations were performed with the resources provided by GENCI-CCRT
(project t2013056929).


%

\end{document}